\documentclass[floatfix,aps,pra,twocolumn,showpacs,amsmath,preprintnumbers,nofootinbib,secnumarabic,superscriptaddress]{revtex4} 

\usepackage{epsfig,verbatim}
\usepackage{amssymb}
\usepackage{graphicx}
\usepackage{color}

\usepackage{braket}
\usepackage{bm}
\usepackage{url}
\usepackage{setspace}
\usepackage[section]{placeins}
\usepackage{appendix}
\usepackage{afterpage}

\def\showeqnlabel{y}    
\def\shownts{y}

\def\bfl{\begin{flushleft}}
\def\efl{\end{flushleft}}
\def\bfr{\begin{flushright}}
\def\efr{\end{flushright}}
\def\bc{\begin{center}}
\def\ec{\end{center}}
\def\be{\begin{equation}}
\def\ee{\end{equation}}
\def\ba{\begin{eqnarray}}
\def\ea{\end{eqnarray}}
\def\baa#1{\begin{array}{#1}}
\def\eaa{\end{array}}
\def\bw{\begin{widetext}}
\def\ew{\end{widetext}}
\def\nn{\nonumber }

\def\bit{\begin{itemize}}
\def\eit{\end{itemize}}

\newcommand{\ola}[1]{\overleftarrow{#1}}
\newcommand{\ora}[1]{\overrightarrow{#1}}

\newcommand{\eqnlabel}[1]{\if y\showeqnlabel
\;\textcolor[rgb]{0.2,0.6,0.2}{\text{#1}}\fi}

\newcommand{\nts}[1]{\if y\shownts \;\colorbox[rgb]{1,0.1,0.1}{NOTE}
\textcolor[rgb]{1,0.1,0.1}{#1}\fi}

\begin{document}

\title{Non-Hamiltonian modeling of squeezing and
thermal disorder in driven oscillators}

\author{Sashwin Sewran}
\email{sash.sewran@gmail.com}
\affiliation{
School of Chemistry and Physics, University of KwaZulu-Natal in Pietermaritzburg, 
Private Bag X01, Scottsville 3209, South Africa} 

\author{Konstantin G. Zloshchastiev}
\email{k.g.zloschastiev@gmail.com}
\affiliation{
School of Chemistry and Physics, University of KwaZulu-Natal in Pietermaritzburg, 
Private Bag X01, Scottsville 3209, South Africa}

\author{Alessandro Sergi}
\email{sergi@ukzn.ac.za}
\affiliation{
School of Chemistry and Physics, University of KwaZulu-Natal in Pietermaritzburg, 
Private Bag X01, Scottsville 3209, South Africa}
\affiliation{KwaZulu-Natal Node,
National Institute for Theoretical Physics (NITheP), South Africa}

\begin{abstract}
Recently, model systems with
quadratic Hamiltonians and time-dependent interactions were studied
by Briegel and Popescu and by Galve \emph{et al}
in order to consider the possibility of both quantum refrigeration in enzymes
[Proc. R. Soc. {\bf 469} 20110290 (2013)]
and entanglement in the high temperature limit [Phys. Rev. Lett. {\bf 105}
180501 (2010); Phys. Rev. A {\bf 81} 062117 (2010)].
Following this line of research,
we studied a model comprising two quantum
harmonic oscillators driven by a time-dependent harmonic coupling.
Such a system was embedded in a thermal bath represented
in two different ways. In one case, the bath was composed of
a finite but great number of independent harmonic oscillators
with an Ohmic spectral density.
In the other case, the bath was more efficiently
defined in terms of a single oscillator
coupled to a non-Hamiltonian thermostat.
In both cases, we simulated the effect of the thermal disorder
on the generation of the squeezed states in the two-oscillators
relevant system.
We found that, in our model, the thermal disorder of the bath determines
the presence of a threshold temperature, 
for the generation of squeezed states, equal to $T=311.13$ K.
Such a threshold is estimated to be within temperatures where
chemical reactions and biological activity comfortably take place.
\end{abstract}

\date{\footnotesize Received: 26 November 2014 [arXiv]}

\pacs{42.50.Dv, 05.30.-d, 07.05.Tp, 74.40.Gh\\
Mathematics Subject Classification (2000): 82C10, 81S30, 00A72, 37N20\\
Keywords: Quantum State Squeezing, Thermal Disorder, Quantum Dynamics, Wigner Function}

\maketitle

%%%%%%%%%%%%%%%%%%%%%%%%%%%%%%%%%%

\section{Introduction}

The idea that quantum mechanics plays a fundamental role in the 
functioning of living matter
is both old and illustrious~\cite{whatislife}.
This concept has been recently revived both by researchers in the field
of quantum information theory~\cite{qinfo}
and by the steady accumulation of experimental evidence supporting
the relevance of high-temperature quantum effects 
in organic molecules and biological systems~\cite{engel,collini,pani,fle11}.
Moreover, it has been suggested that time-dependent couplings might lead
to intra-molecular refrigeration in enzymes \cite{bri13} so that 
low temperatures, where the magnitude of quantum effects is greater,
can be reached with a well-defined mechanism.

What is more relevant to the present work is that non-equilibrium conditions
might enhance quantum dynamical effects 
in biological and condensed matter systems.
For instance, quantum resonances have been found to raise
the critical temperature of superfluid
condensation by means of a mechanism similar to that provided by 
the Feshbach resonance 
in ultra cold gases \cite{poc09}. By analogy, resonances
have also been proposed to be relevant in high-temperature 
superconductors \cite{vbb97} and in living matter \cite{chi10}.
Moreover, recent theoretical studies on model systems driven out of
equilibrium \cite{EntangHighT,EntangHighT2,gue12,pachon}
have supported the persistence of quantum entanglement~\cite{amico,horodecki}
at high temperatures.

The usual approach to the dynamics of open quantum systems is realized
through master equations~\cite{lecture_notes}
or path integrals~\cite{weiss}, for example.
Non-harmonic and non-Markovian dynamics prove to be a tough
problem when attacked with these theoretical tools.
In the present work, instead, we use the 
Wigner representation~\cite{wigner,DistFunctions,lee}
of quantum mechanics and the generalization
of techniques originally stemming from
molecular dynamics simulations~\cite{Liquids,UnderstandMolSim}.
In this work, such techniques are employed to investigate the generation
of squeezing~\cite{QOptics,quantumnoise} at high temperature
under non-equilibrium conditions.
Our study is performed on a model of two harmonic oscillators (representing
two modes of an otherwise general condensed matter system)
embedded in a dissipative bath. The oscillators are coupled
in a time-dependent fashion in order to mimic the action of
an external driving (which might also be caused by
some unspecified conformational rearrangement of the bath)
while the dissipative bath has been formulated in two alternative ways
(which provide equivalent results in our simulations).
In the first case, the bath is specified in terms of a finite number
of independent harmonic oscillators with an Ohmic spectral density.
In the second case, the bath is realized through a single harmonic
oscillator coupled to a non-Hamiltonian thermostat (\emph{i.e.},
a Nos\'e-Hoover Chain thermostat~\cite{nhc}).
Such a non-Hamiltonian thermostat is defined in terms
of two free parameters,
$m_{\eta_1}$ and $m_{\eta_2}$, which play the role
of fictitious masses.
We observed that, for the model studied,
the agreement between the results (obtained by means of the two different
representations of the bath) is achieved within the range of values
$0.96\le m_{\eta_1}=m_{\eta_2} \le 1.04$.

It is known that when the environment is formed by a single bath
there can be decoherence-free degrees of freedom~\cite{deco1,deco2,deco3}.
Hence, a single bath can be expected to lead more easily to
the preservation of quantum effects in general.
Nevertheless, there are circumstances in which a single bath
is exactly what is required by the physical situation.
For example, when the relevant system is formed by a localized
mode (in a somewhat small molecule) which is not under
the influence of thermodynamic gradients, the modeling of
the environment by means of a single bath appears to be
physically sound.
In any case, it is worth mentioning that the computational scheme 
presented in this work
can be easily generalized to describe multiple dissipative baths.
Indeed, within a partial Wigner representation this has already
been done in Ref.~\cite{ilya}.

Squeezed states have widespread applications, especially in
experiments which are limited 
by quantum noise\cite{quantumnoise}.
The control of quantum fluctuations can be used to limit the
sensitivity in quantum experiments. 
Some of these applications can be found in condensed
matter \cite{phonon_sqz,condense_squeeze,condense_squeeze2}, 
in spectroscopy \cite{spec_squeeze},
in quantum information \cite{qi_squeeze} and in gravitational 
wave detection \cite{grav_squeeze}. 
Very often, squeezed states are the concern of quantum optics where
the quadratic degrees of freedom are photons.
However, in a condensed matter system, one still has 
quadratic degrees of freedom, given by phonons, 
so that the theory of squeezing in quantum optics
can be translated to quantum condensed matter systems.

If squeezing could be present at high temperatures within biological macromolecules,
one could speculate about its role in the passage of a substrate through
an ion channel: the reduction (squeezing) of the amplitude of the fluctuation
of the substrate's position might favor its passage through the channel.
The squeezing of the fluctuations of only specific molecules (selectivity)
might arise from the resonance between the substrate's molecular vibrations and
the phonons characterizing the channel (in analogy with what has been proposed
in Ref.~\cite{odor} concerning odor sensing).
However, the above example will only be left as speculative motivation
driving the present work, which is solely concerned with the modeling of
thermal disorder in the squeezing of molecular vibrations.
To this end, we adopt the Wigner representation of
quantum mechanics~\cite{wigner,DistFunctions,lee}
and simulate numerically
the quantum non-equilibrium statistics of our model.
For our quadratic Hamiltonian, quantum dynamics can be
represented in terms of the classical evolution of a swarm of trajectories
with a quantum statistical weight, which is
determined by the chosen thermal initial conditions.
Quantum averages are, therefore, calculated in phase space, as in standard
molecular dynamics simulations~\cite{Liquids,UnderstandMolSim}.
The generation of squeezed states is monitored through the 
threshold values
of the average of suitable dynamical properties~\cite{QOptics,quantumnoise}.
The dependence of the generated amount of squeezing
on the temperature of the environment is investigated.
It is found in our model that there is a temperature threshold 
for squeezed states generation.
The temperature and the time scale at which such a threshold is located
are in the range where the dynamics and chemical reactions
in biological systems occur.

The interest of the this work is twofold.
Firstly, it is a methodological study aiming at verifying the effectiveness
of simulation techniques (based on the Wigner representation of quantum
mechanics) when calculating time-dependent effects in open quantum systems.
At present, such techniques are not commonly used when studying
open quantum systems. However, they promise
a somewhat straightforward extension to non-harmonic couplings
and non-Markovian dynamics.
Secondly, we find that our model, under the conditions adopted  for 
the calculation in the present study, confirms
that quantum squeezing can be present at temperatures
of relevance for biological functioning.

This paper is structured in the following way.
In Sec.~\ref{sec:wigner} we sketch the Wigner representation
of quantum mechanics
and its use in conjunction with temperature control through a 
Nos\'e-Hoover Chain
non-Hamiltonian thermostat.
In Sec.~\ref{sec:model} we introduce our model, together with the different
ways we represent its dissipative environment.
The algorithm for sampling the initial conditions, propagating the
classical-like trajectories (which represent the quantum evolution of the Wigner
function), and the way we monitor the formation of squeezed states in the
simulation are illustrated in Sec.~\ref{sec:algorithm}.
Numerical results are discussed in Sec.~\ref{sec:results}.
Finally, our conclusions and perspectives are presented in Sec.~\ref{sec:conclusions}.

%%%%%%%%%%%%%%%%%%%%%%%%%%%%%%%%%%%%%%%%%%%%%%%%%%%%

\section{Wigner representation}
\label{sec:wigner}

The Wigner function, expressed in the position basis, is 
defined as a specific integral transform of the density 
matrix~\cite{wigner,DistFunctions,lee} $\hat{\rho}(t)$ of the system under study:
\bw
\be\label{eq:Wfunc}
W(r,p,t) = \frac{1}{(2\pi \hbar)^{N_f}}
\int_{-\infty}^{+\infty} d^{N_f}ye^{-\frac{ipy}{\hbar}} 
\Braket{r+\frac{y}{2}|\hat{\rho}(t)|r-\frac{y}{2}} \; ,
\ee
\ew
where $N_f$ is the number of degrees of freedom and a multidimensional notation
is adopted, so that $(r,p,y)$ stands for $(r_i,p_i,y_i)$, with $i=1,...,N_f$.
Using the Wigner representation, quantum statistical averages are calculated as
\begin{eqnarray}
\left< \chi\right> &=& \int_{-\infty}^{+\infty}\int_{-\infty}^{+\infty}  
W(r,p) \chi_W(r,p) d^{N_f}rd^{N_f}p \; ,
\label{avgWig}
\end{eqnarray}
where $\chi_W(r,p)$ is the Wigner representation of the quantum operator $\hat{\chi}$;
such a representation is obtained by considering an integral transform equal to those
in Eq.~(\ref{eq:Wfunc}) but without the pre-factor $(2\pi\hbar)^{-N_f}$.
Since in general
the Wigner function can have negative values because of
quantum interference~\cite{QphysVlad}, it is interpreted as a 
quasi-probability 
distribution function~\cite{wigner,DistFunctions,lee,QphysVlad,Ballentine}.

One of the advantages provided by the use of the Wigner representation
of quantum mechanics is that the equation of motion of the density matrix,
\begin{eqnarray}
\frac{\partial \hat\rho}{\partial t} &=& -\frac{i}{\hbar}
\left[\hat H,\hat\rho\right]\;,
\label{chi_heisen}
\end{eqnarray}
is mapped onto the classical Liouville
equation for $W(q,p,t)$ when the Hamiltonian operator $\hat H$ of
the system is quadratic.
To see this, one can consider the
Hamiltonian operator of system comprising of $N$ harmonic modes:
\begin{equation}
\hat H = \sum_{n=1}^N\left(\frac{1}{2m}\hat P_n^2 + 
\frac{1}{2}m\omega_n^2\hat R_n^2\right) \label{ham} \; .
\end{equation}
Here $\hat P_n$, $\hat R_n$ and $\omega_n$ are the momentum operator, position operator 
and frequency of mode $n$ respectively. For simplicity, each mode is given equal mass $m$.\\
The Wigner representation of the equation of motion~(\ref{chi_heisen})
is, in general,
\be
\frac{\partial W(q,p,t)}{\partial t} =
 \frac{2}{\hbar}
H_W\sin\left[\frac{\hbar}{2}
\left(\frac{\ola \partial }{\partial r}
\frac{\ora \partial}{\partial p}-\frac{\ola \partial }{\partial p}
\frac{\ora \partial}{\partial r}\right)\right]
W(q,p,t)
.
%\ \label{Moyal} \; .
\label{eq:Weqofm}
\ee
The Wigner-transformed Hamiltonian $H_W$ is obtained from the
quantum operator in Eq.~(\ref{ham}) with the substitution $\hat{P}_n\to p_n$,
$\hat{R}_n\to r_n$, for $n=1,...,N$.
However, since the Hamiltonian only contains quadratic terms in both position and momentum,
the Wigner equation of motion~(\ref{eq:Weqofm})
reduces to the classical Liouville equation
\be
 \frac{\partial W(q,p,t)}{\partial t} 
=
\left(\frac{\partial H_W}{\partial r}
\frac{\ora \partial}{\partial p}-\frac{\partial H_W}{\partial p}
\frac{\ora \partial}{\partial r}\right)W(q,p,t).
\label{class_dyn}
\ee
Equation~(\ref{class_dyn}) has a purely classical appearance
whereas all quantum effects arise from the initial conditions. 
When the initial state of the $N$-oscillator system
is positive-definite (as in the case of a thermal state),
Eq.~(\ref{class_dyn}) makes it possible to simulate the quantum dynamics of
a purely harmonic system via classical methods.

In Ref.~\cite{nonHamTherm} it was shown how the quantum evolution
in the Wigner representation can be generalized in order to
control the thermal fluctuations of the phase space coordinates $(r,p)$.
This was achieved upon introducing a generalization of the Moyal bracket~\cite{Moyal_paper}
that extended the Nos\'e-Hoover thermostat~\cite{nose,hoover} to quantum Wigner phase space.
Similarly, it was shown in~\cite{nonHamTherm} how to apply
the so-called Nos\'e-Hoover Chain (NHC) thermostat~\cite{nhc} to Wigner dynamics
in order to achieve a proper temperature control for stiff oscillators.
In the following, we will briefly sketch the theory by specializing it to
harmonic systems. However, since the Wigner NHC method
is not common in the theory of open quantum systems, we provide
a somewhat extended introduction in Appendix~\ref{app:nosewigner}.
In order to introduce the Wigner NHC dynamics for harmonic systems, one can
consider the Wigner-transformed Hamiltonian $H_W$,
introduce four additional fictitious variables and
define an extended Hamiltonian as
\be
H^{\rm NHC} =
H_W + \frac{p_{\eta_1}^2}{2m_{\eta_1}} + 
\frac{p_{\eta_2}^2}{2m_{\eta_2}} + gk_BT_{\rm ext}\eta_1 + k_BT_{\rm ext}\eta_2 
,
\label{eq:H-NHC}
\ee
where $(\eta_1,\eta_2,p_{\eta_1},p_{\eta_2})$ denote the fictitious variables
with masses $m_{\eta_1}$ and $m_{\eta_2}$, respectively.
The symbol $g$ denotes the number of degrees of freedom 
to which the NHC thermostat is attached, $k_B$ is the Boltzmann constant and $T_{\rm ext}$ 
is the absolute temperature of the bath. The phase space point of the extended system 
is defined as $x = \left(r,\eta_1,\eta_2,p,p_{\eta_1},p_{\eta_2}\right)$ .
Introducing the antisymmetric matrix $\mbox{\boldmath $\cal B$}^{\rm NHC}$,
\begin{eqnarray}
\mbox{\boldmath$\cal B$}^{\rm NHC}
&=& \left[ \begin{array}{cccccc}
0 & 0 & 0 & 1 & 0 & 0  \\
0 & 0 & 0 & 0 & 1 & 0  \\
0 & 0 & 0 & 0 & 0 & 1  \\
-1 & 0 & 0 & 0 & -p & 0 \\
0 & -1 & 0 & p & 0 & -p_{\eta_1} \\
0 & 0& -1 & 0 & p_{\eta_1} & 0 \\
\end{array} \right] \;,
\label{eq:B-NHC}
\end{eqnarray}
it is possible to express the NHC equations of motion 
as~\cite{nheom,nh_esm,geometry}
\begin{eqnarray}
\dot x_j &=& \mbox{\boldmath$\cal B$}^{\rm NHC}_{jk} \frac{\partial H^{\rm NHC}}{\partial x_k} \; ,
\label{eq:WNHCeqofm}
\end{eqnarray}
where the Einstein notation of summing over repeated indices has been used.
Hence, as shown in~\cite{nonHamTherm},
in order to achieve temperature control,
Eq.~(\ref{class_dyn}) must be replaced by
\begin{equation}
\frac{\partial W(x,t)}{\partial t}
=-\frac{\partial W(x,t)}{\partial x_j}{\cal B}_{jk}^{\rm NHC}
\frac{\partial H^{\rm NHC}}{\partial x_k}
\label{eq:Weqofm-extended}
\end{equation}
in the extended phase space.
Equation~(\ref{eq:Weqofm-extended}) is called the Wigner NHC equation
of motion. It also contains 
quantum-corrections over the fictitious NHC variables
$(\eta_1,\eta_2,p_{\eta_1},p_{\eta_2})$.
However, its was shown in Ref.~\cite{nonHamTherm}
that a classical limit on the dynamics of such variables
can be taken in order to avoid spurious quantum effects
and represent only the thermal fluctuation of the environment.
Further details can be found in Appendix~\ref{app:nosewigner}.

%%%%%%%%%%%%%%%%%%%%%%%%%%%%%%%%%%%%%%%%%%%%%%%%%%%%%%%%%%
\section{Model System}
\label{sec:model}

In this work, we simulated a model comprising a relevant system and
an environment. The relevant system is given by two 
coupled quantum harmonic oscillators. 
The environment was represented in two different ways,
which will be described in Secs.~\ref{sec:ohmic-bath}
and~\ref{sec:nhc}. Here, we first introduce the relevant system.

\begin{figure}[htbt]
\begin{center}
	\epsfig{figure=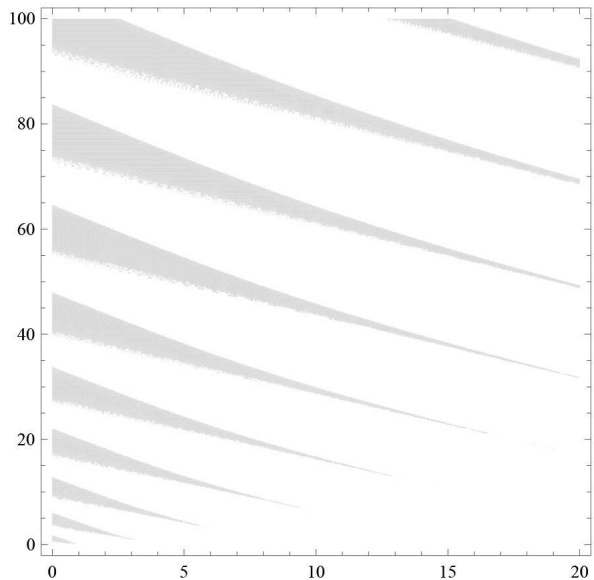,width=  0.9\columnwidth}\end{center}
\caption{Instability region (shaded area) in the parametric space of the 
model (\ref{Ham_2osc}), namely, in terms of values of $(\omega / \omega_d)^2$
(horizontal axis) versus $(\omega_0 / \omega_d)^2$. The values reported on the
axes are in dimensionless units. }
\label{f:instab}
\end{figure}

In the relevant system the coupling between the oscillators is 
oscillatory, time-dependent and quadratic. 
In the Wigner representation, the Hamiltonian of the system is
\begin{equation}
H_{\rm S} = \frac{p_1^2}{2m} + \frac{p_2^2}{2m} 
+ \frac{m \omega^2}{2} \left(q_1^2+q_2^2 \right)
+\frac{m\tilde{\omega}^2(t)}{2}\left(q_2-q_1\right)^2 \;,
\label{Ham_2osc}
\end{equation}
where $\omega$ is the proper frequency of the oscillators,
$\omega = \sqrt{K/m}$, and
$\tilde{\omega}(t)$ is the time-dependent frequency of
the coupling between the oscillators
\be
\tilde{\omega}(t) \equiv \sqrt{\tilde{K}(t) / m} 
= \omega_0 \sin\left(\omega_d t\right)  \;.
\ee
Here $p_1$ and $p_2$ are the momenta of the oscillators, 
$m$ is the mass of both oscillators, $q_1$ and $q_2$ are the 
displacement of the oscillators from their equilibrium positions, 
$K$ is the spring constant of both oscillators, $\omega_0$ is the amplitude 
frequency of the coupling, $\omega_d$ is the driving frequency and $\tilde{K}(t)$ is 
the coupling function between the oscillators. 

In Refs.~\cite{EntangHighT} and~\cite{linearquantum}
analytical solutions to similar models have been found. 
However, our system differs in the time dependence of the 
coupling between the oscillators. 
On a classical level our model can be treated in terms 
of the Mathieu functions and using the Floquet theorem~\cite{Floquet}.
Using the notation  defined in Appendix \ref{s-app}
(and assuming $\omega_c \equiv \omega_d$, where $\omega_c$ is 
the frequency characterizing the spectral density of the
bath introduced in Sec.~\ref{sec:ohmic-bath}), we obtain the
following equations of motion
\ba
&&
\frac{d^2 Q_1'}{d t^{\prime 2}} + \frac{\omega^{2}}{\omega_d^{2}}
Q_1' = 0 \;, \label{eq:eqofmQ1}\\
&&
  \frac{d^2 Q_2'}{d t^{\prime 2}} +
\left[
\frac{\omega^{2} + \omega_0^{2}}{\omega_d^{2}}
-
\frac{\omega_0^{2}}{\omega_d^{2}}
\cos{(2 t')}
\right]
Q_2' = 0 \;, \label{eq:eqofmQ2}
\ea
where $Q_1' = (q_1'+ q_2')/\sqrt{2}$ and $Q_2' = (q_1'-q_2')/\sqrt{2}$
are the dimensionless center-of-mass and relative displacement coordinates,
respectively.
While the solution of Eq.~(\ref{eq:eqofmQ1}) is simply
a linear combination of sine and cosine functions,
Eq.~(\ref{eq:eqofmQ2}) is the Mathieu equation and
possesses more complex features.
In particular, for certain values of its parameters,
it develops dynamical instabilities, see Fig.~\ref{f:instab}.
Such parameters values must be avoided when doing the 
numerical simulations in the quantum case.

\subsection{Ohmic bath}
\label{sec:ohmic-bath}

In order to represent dissipative effects, the relevant system
described by the Hamiltonian $H_{\rm S}$ was coupled, via a bilinear coupling,
to a bath of $N$ independent harmonic oscillators
with an Ohmic spectral density~\cite{leggett}.
The total Hamiltonian is
\begin{equation}
H^{\rm NB}=H_{\rm S}+H_{\rm B}+H_{\rm SB}
\label{NOSC_ham}
\end{equation}
where
\begin{eqnarray}
H_{\rm B}&=&\sum_{j=1}^N \left(\frac{P_{j}^2}{2m_j}
+\frac{1}{2}m_j\Omega_j^2R_j^2\right)\;, \label{eq:HB} \\
H_{\rm SB}&=& -\sum_{\alpha=1}^2\sum_{j=1}^N q_\alpha c_jR_j \; .
\label{eq:HSB}
\end{eqnarray}
The parameters in Eqs.~(\ref{eq:HB}) and~(\ref{eq:HSB}) are defined as
\begin{eqnarray}
\Omega_j &=& -\omega_c \ln{\left(1-j\frac{\bar\omega_0}{\omega_c}\right)}
\label{wj}\\
\bar\omega_0 &=&
\frac{\omega_c}{N}\left[1-
\exp{\left(-\frac{\omega_{\rm max}}{\omega_c}\right)}\right]
\label{eq:baromega0}
\\
c_j &=& \sqrt{\xi \hbar \bar\omega_0 m_j}\Omega_j  \; .  \label{cj}
\label{Hbath}
\end{eqnarray}
The frequency $\omega_{\rm max}$ in Eq.~(\ref{eq:baromega0})
is a cut-off frequency used in the numerical representation
of the spectral density. The value of $\omega_{\rm max}$
used in the calculations reported in this work is given in
Sec.~\ref{sec:algorithm}.
Each oscillator in the bath has a different frequency, $\Omega_j$.
The definition of $\Omega_j$, $\bar\omega_0$ and $c_j$ is chosen in such
a way to represent 
an infinite bath of oscillators with Ohmic spectral density~\cite{leggett}
in terms of discrete mode of oscillations~\cite{makri,linear1,linear2}.
The parameters $\xi$ and $\omega_c$ characterize the
spectral density of the bath.  The Kondo parameter, $\xi$,
is a measure of the strength of the coupling between the relevant system
and the bath.

\begin{figure}[htbt]
\begin{center}
	\epsfig{figure=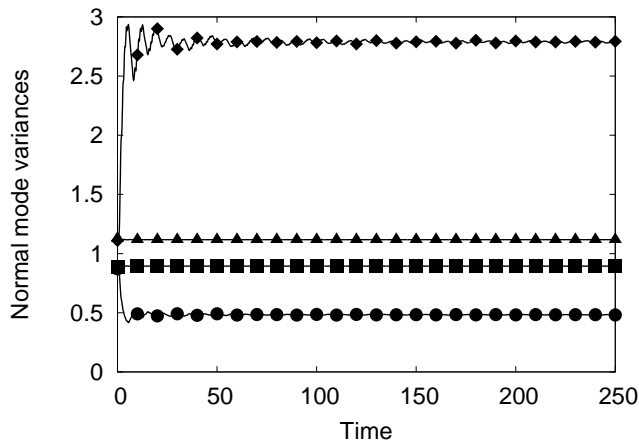,width=  1.01\columnwidth}\end{center}
\caption{Variance of the position and momentum coordinate of each normal mode, 
where the relevant 
system is attached to a harmonic bath. The shaded square, circle, 
triangle and diamond points
represent the variances of $R_1$, $R_2$, $P_1$ and $P_2$ respectively.
Solid lines, connecting the numerically calculated points,
have been drawn to guide the eye. 
Dimensionless parameters used in this simulation:
$m=1.0$, $K=1.25$, $\omega_0=2.50$, $\omega_d=0.45$,
$\xi=0.007$ and $T_{\rm ext}=1.0$. The values reported on the
axes are also in dimensionless units.}
\label{2osc_varall}
\end{figure}

\subsection{NHC representation of the bath}
\label{sec:nhc}

We adopted a second technique to represent
the dissipative environment in which the relevant driven system
is embedded.
In particular, we considered a single oscillator bilinearly coupled
to the relevant system and we thermalized it
by means of a Nos\'e-Hoover Chain~\cite{nhc,nonHamTherm}.
Such a technique (and similar ones~\cite{dlamini,b3}) allows one
to reduce drastically
the computational time by representing the thermal environment
with a minimal number of degrees of freedom. 
In this case, the total Hamiltonian is
\begin{eqnarray}
H^{\rm NHC} &=& H_{\rm S} + H_{\rm B}^1 + H_{\rm SB}^1 
+H^{\rm TH} \;, \label{NHC_ham}
\\
H_{\rm B}^1&=&\frac{P_1^2}{2m_1}+\frac{1}{2}M\Omega_1^2R_1^2 \;,
\\
H_{\rm SB}^1&=&- c_1R_1\left(q_1+q_2\right) \;,
\\
H^{\rm TH}&=&\frac{P_{\eta_1}^2}{2m_{\eta_1}} + \frac{P_{\eta_2}^2}{2m_{\eta_2}} 
+ k_BT_{\rm ext}\eta_1+k_BT_{\rm ext}\eta_2 \; .
\end{eqnarray}
Here
$P_1$ and $R_1$ are the phase space variables of the bath oscillator having mass $m_1$ 
and frequency $\Omega_1$. 
The bath and driven system are bilinearly coupled.
The fictitious Nos\'e variables are indicated by
$\eta_1$ and $\eta_2$ while $P_{\eta_1}$ and $P_{\eta_2}$ are 
their associate momenta.
The fictitious Nos\'e variables have masses
$m_{\eta_1}$ and $m_{\eta_2}$, respectively.
The symbol $k_B$ denotes 
the Boltzmann constant while $T_{\rm ext}$ 
indicates the absolute temperature of the bath. 
As explained with more detail in Appendix~\ref{app:nosewigner},
the coupling to the fictitious thermostat variables
$(\eta_1,\eta_2,p_{\eta_1},p_{\eta_2})$ is realized
through the non-Hamiltonian equation of motion.
In the classical case, such equations are 
written in compact form in Eq.~(\ref{eq:WNHCeqofm})
or in explicit form in Eqs.~(\ref{eq:eq-nhc1}-\ref{eq:eq-nhc6}).
In the quantum case, the coupling is given
through Eq.~(\ref{eq:Weqofm-extended}).
The quantum-classical approximation of Eq.~(\ref{eq:Weqofm-extended}),
which suppresses the spurious quantum effects over the
fictitious NHC variables, is instead given in Eq.~(\ref{eq:qcnosewig}).
Equation~(\ref{eq:qcnosewig}) is the one used for the NHC
representation of the bath in this work.

\begin{figure}[htbt]
\begin{center}
	\epsfig{figure=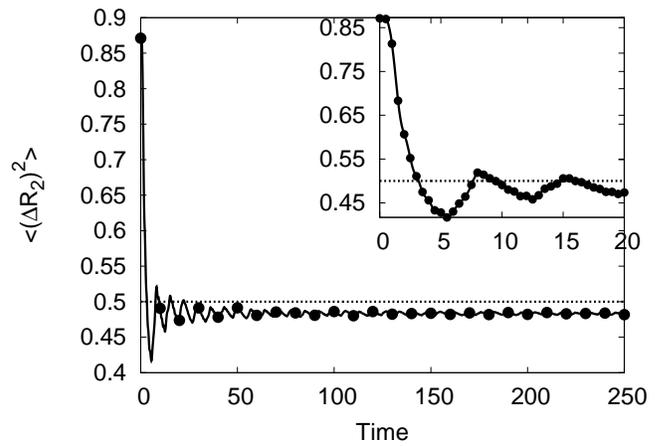,width=  1.01\columnwidth}\end{center}
\caption{Variance of the position coordinate of normal mode 2, $R_2$, 
where the relevant system
is attached to a harmonic bath. 
A horizontal line at $\langle(\Delta R_2)^2\rangle$=0.5
shows the theoretical threshold value for squeezed state generation.
The inset shows the short time simulation of the variance of $R_2$, 
represented by solid circles. 
A solid line, connecting the numerically calculated points, 
has been drawn to guide the eye.
Dimensionless parameters used in this simulation:
$m=1.0$, $K=1.25$, $\omega_0=2.50$, $\omega_d=0.45$,
$\xi=0.007$ and $T_{\rm ext}=1.0$. The values reported on the
axes are also in dimensionless units.
}
\label{2osc_varR2}
\end{figure}

\section{Simulation details}
\label{sec:algorithm}

The algorithm used to integrate the equations of motion in all our simulations
is based
on the symmetric Trotter factorization of the 
propagator~\cite{reversibleintegrators,reverse_int}. 
When we considered the NHC thermostat to represent the thermal bath,
we also incorporated the Yoshida scheme~\cite{yoshida} with 
three iterations and 
a multiple time-step procedure with three iterations,
following the approach of Ref.~\cite{reversibleintegrators}.
In the simulations, we set as initial conditions
for the NHC variables $\eta_1=0$, $\eta_2=0$, $p_{\eta_1}=0$ 
and $p_{\eta_2}=1.0$.

At $t=0$ it is assumed that the system is at thermal equilibrium
with no time-dependent driving. The driving acts for $t>0$.
In this case, the Wigner function of the total system is positive definite
and can be represented as a collection of points that are propagated
according to Eq.~(\ref{class_dyn}), when the bath is represented by
means of $N$ oscillators with an Ohmic spectral density,
or according to Eq.~(\ref{eq:Weqofm-extended}), when using
the NHC bath.

In order to
sample the initial configuration of the relevant system,
it is useful to introduce normal coordinates~\cite{Goldstein}: 
\begin{eqnarray}
\tilde{q}_1 &=& \frac{1}{\sqrt{2}}\left(q_1+q_2\right)\;, \label{eq:norm1}\\
\tilde{q}_2 &=& \frac{1}{\sqrt{2}}\left(q_1-q_2\right) \;, \\
\tilde{p}_1 &=& \frac{1}{\sqrt{2}}\left(p_1+p_2\right) \;,\\
\tilde{p}_2 &=& \frac{1}{\sqrt{2}}\left(p_1-p_2\right) \label{eq:norm4} \; ,
\end{eqnarray}
so that the Hamiltonian $H_{\rm S}$ in Eq.~(\ref{Ham_2osc}) can be written as 
\begin{eqnarray}
H^{\rm NM} &=& \sum_{k=1}^2 \left(\frac{\tilde{p}_k^2}{2m}
+\frac{1}{2}m\omega_k^2\tilde{q}_k^2\right) \; ,
\label{H_n}
\end{eqnarray}
where $\omega_k$ ($k=1,2$) are the normal mode frequencies.
The symbol
$\tilde{q}_1$ represents the motion of the centre of mass of the system,
while
$\tilde{q}_2$ represents the relative displacements of the oscillators.
The normal mode frequencies of each mode are 
\begin{eqnarray}
\omega_1 &=& \sqrt{\frac{K}{m}} \;,
\label{w1}\\
\omega_2(t) &=&\sqrt{\frac{K+2\tilde{K}(t)}{m}} \;. \label{w2}
\end{eqnarray}
The initial conditions of the system are sampled from the 
Wigner function~\cite{DistFunctions}:
\be
W_{\rm S}
=
\prod_{k=1}^2  \frac{1}{\pi\hbar}\tanh
\left(\frac{\hbar\omega_k}{2}\beta\right)
\exp\left(-\frac{\tilde{p}_k^2}{2\sigma_{\tilde{p}_k}^2}\right)
\exp\left(-\frac{\tilde{q}_k^2}{2\sigma_{\tilde{q}_k}^2}\right)
\label{gaussians}
\ee
where
\begin{eqnarray}
\sigma_{\tilde{p}_k} &=&
\left[\frac{2}{\hbar m\omega_k}\tanh
\left(\frac{\hbar\omega_k}{2}\beta\right)\right]^{-\frac{1}{2}} 
\;, \label{sigmaP}\\
\sigma_{\tilde{q}_k} &=& \left[\frac{2m\omega_k}{\hbar}\tanh
\left(\frac{\hbar\omega_k}{2}\beta\right)\right]^{-\frac{1}{2}} 
\label{sigmaR} \; .
\end{eqnarray}

In the high temperature limit $\left(T\rightarrow\infty,\beta\rightarrow0\right)$, 
the Wigner distribution function in Eq.~(\ref{gaussians}) reduces to the classical
canonical distribution function, $Z^{-1}\exp[-\beta H_{\rm S}]$.
Hence, simply by changing the sampling of the
initial conditions of the system, we can study the difference
between the classical and the quantum behavior of the system.

At $t=0$, the Ohmic bath is also assumed to be at thermal equilibrium
with initial Wigner function equal to
\be
W_{\rm B}
=
\prod_{k=1}^N  \frac{1}{\pi\hbar}\tanh
\left(\frac{\hbar\Omega_k}{2}\beta\right)
\exp\left(-\frac{P_k^2}{2\sigma_{P_k}^2}\right)
\exp\left(-\frac{R_k^2}{2\sigma_{R_k}^2}\right)
\label{eq:wig-ohmic}
\ee
where
\begin{eqnarray}
\sigma_{P_k} &=& \left[\frac{2}{\hbar m \Omega_k}\tanh
\left(\frac{\hbar\Omega_k}{2}\beta\right)\right]^{-\frac{1}{2}} 
\label{eq:sigmaP-ohm}\\
\sigma_{R_k} &=& \left[\frac{2m\Omega_k}{\hbar}\tanh
\left(\frac{\hbar\Omega_k}{2}\beta\right)\right]^{-\frac{1}{2}} 
\label{eq:sigmaR-ohm} \; .
\end{eqnarray}
So that the initial Wigner function for the total system is
$W=W_{S}\times W_{\rm B}$.

When using the NHC representation of the bath, $W_{\rm B}$ in 
Eq.~(\ref{eq:wig-ohmic}) reduces to $W_{\rm B}^1$ (which is obtained considering $N=1$)
while the initial condition of the NHC fictitious variables are taken
as $\prod_{n=1}^2\delta(p_{\eta_n}-p_{\eta_n}^0)\delta(\eta_n-\eta_n^0)$,
where $(\eta_n^0,p_{\eta_n}^0)$, $n=1,2$, are some arbitrary fixed values.
In such a case, the total initial Wigner function is
$W=W_{\rm S}\times W_{\rm B}^1 \times \prod_{n=1}^2\delta(p_{\eta_n}-p_{\eta_n}^0)\delta(\eta_n-\eta_n^0)$.

Considering two arbitrary quantum operators, $\hat a$ and $\hat b$, 
satisfying the commutation relation $[\hat a, \hat b] = i\hat c$, it is known
that there is a squeezed state  if~\cite{QOptics}
\begin{equation}
 \left<\left(\Delta a_{\rm W}\right)^2\right> <
\frac{1}{2}\left|\left< c_{\rm W}\right>\right|
\ \mathrm{or} \
\left<\left(\Delta b_{\rm W}\right)^2\right>
< \frac{1}{2}\left|\left<c_{\rm W}\right>\right| \; , 
\end{equation}
where $a_{\rm W}$, $b_{\rm W}$ and $c_{\rm W}$ are the Wigner
representation of $\hat a$, $\hat b$ and $\hat c$, respectively,
and $\Delta a_{\rm W}=a_W-\langle a_{\rm W}\rangle$.
In the case the normal mode coordinates, Eqs.~(\ref{eq:norm1}-\ref{eq:norm4}),
their commutation relations are
$\left[\hat{\tilde{q}}_j,\hat{\tilde{p}}_k\right]=i\hbar\delta_{jk}$, $j,k=1,2$.
In dimensionless coordinates, the conditions of squeezing can be written as
\begin{eqnarray}
\left<\left(\Delta \tilde{q}_k\right)^2\right> < \frac{1}{2}
\quad&\mathrm{or}&\quad 
\left<\left(\Delta \tilde{p}_k\right)^2\right> < \frac{1}{2} \; . 
\label{sqz_thresh}
\end{eqnarray}
Equation~(\ref{sqz_thresh}) provides a threshold for state squeezing.
Upon defining $\chi_k=\tilde{q}_k^2-\langle\tilde{q}_k\rangle^2$ or
$\chi_k=\tilde{p}_k^2-\langle\tilde{p}_k\rangle^2$ for $k=1,2$,
one can use Eq.~(\ref{avgWig}) in order to assess state squeezing.
The squeezed states can be visualized upon constructing
the marginal distribution functions of each normal mode from the numerical
evolution of the total Wigner function.

\section{Numerical Results}
\label{sec:results}

In all simulations we considered an integration time step $dt=0.01$,
a number of molecular dynamics steps $N_{\rm S}= 25 000$, 
and a number of Monte Carlo steps $N_{\rm MC}=10 000$.
Unless stated otherwise, the results are reported in dimensionless coordinates 
and scaled units. 
We performed simulations considering three different cases.
The first concerns the study of the driven oscillators without
the coupling to an external bath.
The second deals with the driven oscillators bi-linearly coupled to 
an Ohmic bath.
The third concerns the driven oscillators coupled to a
dissipative bath constituted by a single harmonic oscillator
thermalized though a Nos\'e-Hoover Chain.
We did not find any appreciable numerical difference
between the results obtained with the Ohmic bath or with
the single harmonic oscillator thermalized though a NHC thermostat.

In order to check our calculation scheme,
we ran a series of simulations without taking into account dissipative effects.
Instead, we focused on the dynamics of the two coupled
oscillators with the Hamiltonian given in Eq.~(\ref{Ham_2osc}).
The stability of the numerical algorithm was tested by calculating
the average value of the energy when considering $\tilde{\omega}=\omega_0$
(which amounts to switching off the time-dependent 
coupling) in Eq.~(\ref{Ham_2osc}).
Such an average value was found to be conserved in
one part over ten thousand.
Upon reintroducing the time-dependent frequency in Eq.~(\ref{Ham_2osc}),
we also verified that a squeezed state is generated.
Starting from a thermal state,
the variance of the position and momentum of each normal mode coordinate
was calculated. The variance of the position of normal mode 2 was found to 
be below the threshold for squeezing 
while the variance of the momentum of normal mode 2 increases
simultaneously
(in agreement with the Heisenberg uncertainty principle).
This indicated that
a squeezed state for the position coordinate of normal mode 2 had been generated.

\begin{figure}[htbt]
\begin{center}
	\epsfig{figure=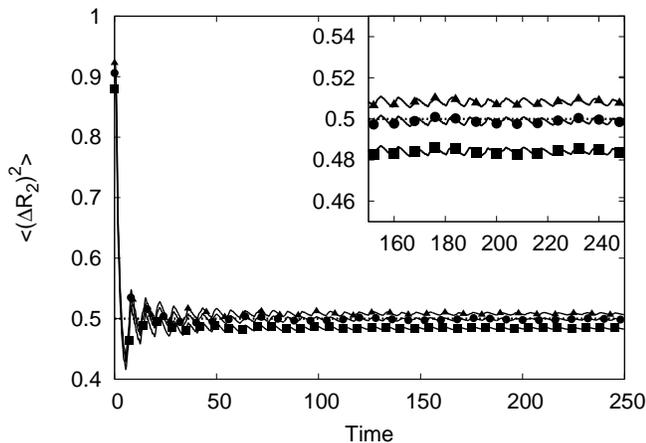,width=  1.01\columnwidth}\end{center}
\caption{Variance of normal mode coordinate $R_2$,
with three different temperatures of the bath.
The square, circle, and triangle points are
for temperatures of the bath 
1.0, 1.037 and 1.06,  respectively.
The inset shows the same curves in the long-time region
in order to better appreciate their differences.
Solid lines, connecting the numerically obtained points,
have been drawn to guide the eye. 
Moreover, a horizontal line at $\langle(\Delta R_2)^2\rangle$=0.5
shows the theoretical threshold value for squeezed state generation.
Dimensionless parameters used in this simulation:
$m=1.0$, $K=1.25$, $\omega_0=2.50$, $\omega_d=0.45$, and
$\xi=0.007$. The values reported on the
axes are also in dimensionless units.}  
\label{threshold}
\end{figure}

In order to account for dissipative effects and study the influence 
of a thermal bath on squeezed state generation, we used two methods
that, as expected~\cite{dlamini,b3},
provided numerically indistinguishable results.
In the first approach, we used a bath of $N=200$ 
harmonic oscillators, bi-linearly coupled to the system of driven 
oscillators. It has been shown that such a discrete representation
of the Ohmic bath is in agreement with
linear response theory~\cite{linear1,linear2}.
The Hamiltonian of this system is given in Eq. (\ref{NOSC_ham}).
In the second approach, we represented the bath by means of
a single oscillator coupled to a NHC thermostat.
In this second case, the Hamiltonian of the system
is given in Eq.~(\ref{NHC_ham}). The values of the fictitious masses
$m_{\eta_1}$ and $m_{\eta_2}$ control the dynamics of the
thermostat, which in turn simulates the thermal bath.
Such masses are tunable parameters that can be adjusted
in order to obtain an optimal agreement between the calculations with
$N=200$ and no thermostat and those with $N=1$ with the thermostat.
In the calculations reported in this work, we set $m_{\eta_1}=m_{\eta_2}=1.0$
(in dimensionless units). We also observed that, for the model studied,
the agreement between the results (obtained by means of the two different
representations of the bath) is achieved within the range of values
$0.96\le m_{\eta_1}=m_{\eta_2} \le 1.04$ of the fictitious masses.

All calculations were performed using 
$\omega_c=1.0$  and $\omega_{\rm max}=3.0$.
Both weak and strong 
coupling strengths to the bath ($\xi=0.007, 0.3$) were studied.
We observed that the coupling strengths influenced the behavior of
normal mode 1 while normal mode 2 remained almost invariant.
Figure~\ref{2osc_varall} shows the variance of each normal mode coordinate. 
The variance of the coordinates
of normal mode 1 maintain a constant value even when the driving
is switched on between the oscillators. 
However, we observe a decrease in the variance of the position 
and an increase in the variance of the momentum of normal mode 2, 
respectively. 
This arises from the
form of the normal mode frequencies in Eqs. (\ref{w1}) and (\ref{w2}):
only the frequency of normal mode 2 is time-dependent,
while the frequency of normal mode 1 has a constant value.
In Fig. \ref{2osc_varR2} we have plotted the theoretical threshold
for squeezing as defined in Eq.~(\ref{sqz_thresh}).
The variance of the position of normal mode 2 is clearly 
below such a threshold.
This shows that the driven oscillators can make a transition
from a thermal to a squeezed state, as expected.
The inset in Fig. \ref{2osc_varR2} shows the dynamics over a short time interval
of the variance of the position of normal mode 2.
The variance goes below the squeezing threshold after a time interval $\Delta t=3.2$
and remains below such a threshold for the rest of its time evolution. 
We can conclude that the generation of a squeezed state 
is fast in comparison to the natural dynamics of the system.
At time $t=0$, the marginal distribution functions of 
both normal modes are symmetrical and have a
circular uncertainty domain~\cite{schleich}.
After the driving is turned on, the marginal distribution function of
normal mode 1 maintains its initial features,
while the marginal distribution function 
of normal mode 2 becomes asymmetrical with an elliptical uncertainty domain.

We also studied the influence of the temperature
on the degree of squeezing of the normal modes.
In particular, we have performed simulations for various
values of the temperature ranging from 0.95 to 1.06
(and used an increment of 0.01).
In Figure \ref{threshold},  the results of three different 
bath temperatures 
($T_{\rm ext}$=1.0, 1.037, and 1.06) are shown.
The value of $T_{\rm ext}$=1.037 provides  a reliable estimate
of the threshold temperature
for the creation of squeezed states in our model, \emph{i.e.},
above this temperature no squeezing is produced by the time-dependent
dynamics. As a matter of fact, we verified that for $T_{\rm ext}$=1.038
we did not observe any squeezing.
Below $T_{\rm ext}$=1.037, for the various temperatures calculated,
we have always observed the creation of squeezed states in our model.

\section{Conclusions and perspectives}
\label{sec:conclusions}

We studied the generation of squeezed states induced
by time-dependent quadratic coupling between two harmonic oscillators
embedded in a dissipative environment.
The latter was represented in two different ways that provided
numerically indistinguishable results. In one approach, we used a bath
of $N=200$ harmonic oscillators with an Ohmic spectral density;
in the other approach, we used a single harmonic oscillator
whose temperature was controlled through a Nos\'e-Hoover 
Chain non-Hamiltonian thermostat.
The quantum equations of motion were mapped onto a classical-like
formalism through the Wigner representation
and integrated numerically by means of standard molecular dynamics techniques. 
The systems studied are relevant in order to model the dynamics
of molecular phonons (for example) in condensed matter.

Upon varying the controlled temperature of the bath in our calculations,
we studied the effect of thermal disorder on the generation
of squeezing.
It was found that there is a threshold temperature of the bath
below which squeezing is still present.
In dimensionless coordinates, this temperature is $T_{\rm ext}=1.037$.
The non-Hamiltonian thermostat is defined in terms of two free parameters,
$m_{\eta_1}$ and $m_{\eta_2}$, which play the role
of fictitious masses.
We observed that, for the model studied,
the agreement between the results (obtained by means of the two different
representations of the bath) is achieved within the range of values
$0.96\le m_{\eta_1}=m_{\eta_2} \le 1.04$.

 If we assume a value of the frequency
$\omega_c=3.93 \times 10^{13}$ Hz we can convert to dimensionful coordinates
and find that such a temperature has the value
$T_{\rm ext}=311.13$ K.
According to this, the quantum of excitation $\hbar \omega_c$ assumes
the value of 4.14 $\times 10^{-21}$ J, which is exactly equal to
the value of $k_BT$ at room temperature
($T_{\rm ext}$=300 K corresponding to $T_{\rm ext}=1.0$
in dimensionless coordinates). With the above value of $\omega_c$,
we found two interesting results.
One is that the time spanned by the simulated trajectories
is of the same order of magnitude as that of molecular oscillations
($\approx 10^{12}$ s).
The second is that the frequency of vibration of the relevant oscillators
is of the order of the terahertz,
as it is expected for molecular functional dynamics
in biological systems~\cite{zhang}. 
Hence, our numerical study supports the possibility of having
observable quantum effects at temperatures that are relevant
for biological functions. 
As suggested by various authors~\cite{poc09,EntangHighT,EntangHighT2,gue12,pachon},
such a counterintuitive occurrence is made possible by the non-equilibrium
conditions arising in the dynamics of coupled molecular systems.
In this work and elsewhere~\cite{poc09,EntangHighT,EntangHighT2,gue12,pachon},
such complex situations have been modeled through phononic modes
with a time-dependent coupling.
However, through a suitable modification of the Wigner trajectory method,
the techniques illustrated in this paper promise 
a somewhat straightforward extension to non-harmonic couplings
and non-Markovian dynamics.

%\begin{acknowledgements}

\section*{Acknowledgments} 

A. S. is grateful to Professor Salvatore Savasta for the interesting discussions
that motivated the present work.
This research was supported through the Incentive Funding for
Rated Researchers by the National Research Foundation of South Africa.
Besides, 
S. S. acknowledges a Ph.D. bursary from the National Institute of Theoretical
Physics (NITheP) in South Africa.

%\end{acknowledgements}

%%%%%%%%%%%%%%%%%%%%%%%%%%%%%%%%%%%%%%%%%%%%%%%%%%

\appendix 

\section{Wigner NHC equations of motion}\label{app:nosewigner}

Let us consider the Hamiltonian in Eq.~(\ref{eq:H-NHC}) and
the antisymmetric matrix in Eq.~(\ref{eq:B-NHC}).
The non-Hamiltonian equations on motion~(\ref{eq:WNHCeqofm})
are written explicitly as
\begin{eqnarray}
\dot{r}&=&\frac{p}{m}\label{eq:eq-nhc1}\;,\\
\dot{\eta_1}&=&\frac{p_{\eta_1}}{m_{\eta_1}}\;,\\
\dot{\eta_2}&=&\frac{p_{\eta_2}}{m_{\eta_2}}\;,\\
\dot{p}&=&-\frac{\partial V}{\partial r}
-\frac{p_{\eta_1}}{m_{\eta_1}}p\label{eq:eq-nhc3}\;,\\
\dot{p}_{\eta_1}&=&\frac{p^2}{m}-gk_BT
-\frac{p_{\eta_2}}{m_{\eta_2}}p_{\eta_1}
\;, \label{eq:eq-nhc4}\\
\dot{p}_{\eta_2}&=&\frac{p_{\eta_1}^2}{m}-k_BT\;.
\label{eq:eq-nhc6}
\end{eqnarray}
The Liouville operator for NHC dynamics is
\begin{eqnarray}
&&
L^{\rm NHC}=
{\cal B}_{ij}^{\rm NHC}\frac{\partial H^{\rm NHC}}{\partial x_i}
\frac{\partial}{\partial x_i}
\nn\\&&\qquad \ \ \,
=
\frac{p}{m}\frac{\partial}{\partial r}
+\frac{p_{\eta_1}}{m_{\eta_1}}\frac{\partial}{\partial \eta_1}
+\frac{p_{\eta_2}}{m_{\eta_2}}\frac{\partial}{\partial \eta_2}
\nn
%\\
\ea
\ba
&&\qquad \ \ \
+\left(-\frac{\partial V}{\partial r}
-\frac{p_{\eta_1}}{m_{\eta_1}}p\right)\frac{\partial}{\partial p}
\nonumber\\&&\qquad \ \ \
+\left(\frac{p^2}{m}-gk_BT\right)\frac{\partial}{\partial p_{\eta}}
\nn\\&&\qquad \ \ \
+\left(\frac{p_{\eta_1}^2}{m}-k_BT\right)\frac{\partial}{\partial p_{\eta_2}}\;.
\label{eq:noseliouvilleop}
\end{eqnarray}
The above equations allow one to define NHC dynamics in
classical phase space.
The coupling between the thermostat momentum $p_{\eta_1}$
and the physical coordinates $p$ is not realized through
the extended
Hamiltonian in Eq.~(\ref{eq:H-NHC}). Instead, it is achieved
through Eq.~(\ref{eq:eq-nhc3}).
Under the assumption of ergodicity,
it can be proven that the NHC equations of 
motion~(\ref{eq:eq-nhc1}-\ref{eq:eq-nhc6})
generate the canonical distribution function for the
physical coordinates $(r,p)$~\cite{nhc,nheom}.

As originally explained in Ref.~\cite{nonHamTherm},
the matrix form of the generalized Wigner bracket given 
in Eq.~\ref{eq:Weqofm-extended} can be used to
define NHC equations of motion in quantum phase space.
Defining the phase space compressibility as
\begin{equation}
\kappa=(\partial_jB_{ji}^{\rm NHC})\partial_iH^{\rm NHC}\;,
\end{equation}
the Wigner NHC equation can be written as
\bw
\begin{eqnarray}
\partial_t  W
&=&-iL^{\rm NHC}  W
-\kappa W
%\nonumber\\&+&
+
\sum_{n=3,5,7,\ldots}\frac{1}{n!}\left(\frac{i\hbar}{2}\right)^{n-1}
%\times\nn\\&& \times
H^{\rm NHC}\left[\overleftarrow{\partial_i}{\cal B}_{ij}^{\rm NHC}
\overrightarrow{\partial_j} +
\overleftarrow{\partial_i}(\partial_j{\cal B}_{ij}^{\rm N})
\right]^nf_W
, \label{eq:fullqnosewig}
\end{eqnarray}
%\ew
where the Nos\'e Liouville operator is defined as in
Eq.~(\ref{eq:noseliouvilleop}) 
and $W=W(r,p,\eta_1,\eta_2,p_{\eta_1},p_{\eta_2},t)$
is the Wigner distribution function in the extended
NHC quantum phase space.
To zeroth order in $\hbar$ the Nos\'e-Wigner equations of motion
coincide with the classical equations of motion. Higher powers of
$\hbar$ provide the quantum corrections to the dynamics. 
The quantum correction terms were considered in more detail
in Ref.~\cite{nonHamTherm}.
However, one is not really interested in the quantum behavior
of the fictitious variables: they are there only to enforce
the canonical distribution and represent a thermal environment.
Moreover, the mass $m_{\eta_1}$ is typically taken to be
much greater than $m$ in order to not modify the dynamical
properties of the system. As a result, one finds
a small expansion parameter 
$\mu=\sqrt{m/m_{\eta_1}}<< 1$ that can be used to take
the classical limit over the $(\eta_1,\eta_2,p_{\eta_1},p_{\eta_2})$
fictitious variables.
In such a way, in place of the full quantum 
equation~(\ref{eq:fullqnosewig})
one obtains
%\bw
\begin{eqnarray}
\partial_t  W
&=&-\left(iL^{\rm NHC}+\kappa\right)  W
%\nn\\&+&\!
+
\sum_{n=3,5,7,\ldots}\frac{1}{n!}\left(\frac{i\hbar}{2}\right)^{n-1}\!
V
\left(\overleftarrow{\partial}_r \overrightarrow{\partial_p}\right)^n  W
.
\label{eq:qcnosewig}
\end{eqnarray}
Equation~(\ref{eq:qcnosewig}) defines a quantum-classical NHC dynamics
according to which the $(r,p)$ coordinates are evolved quantum-mechanically
while the $(\eta_1,\eta_2,p_{\eta_1},p_{\eta_2})$ are evolved classically.
As proven in Ref.~\cite{nonHamTherm}, 
the weak coupling between the two sets of coordinates generates
a canonical distribution function to zero order in $\hbar$.
\ew

\section{Converting equations to dimensionless form}\label{s-app}

It is convenient to introduce the following dimensionless variables:
\ba
&&
q^\prime_i = q_i \sqrt{ \frac{m\omega_c}{\hbar}}, \
p^\prime_i =  \frac{p_i}{\sqrt{m E_c}}, \\&&
R^\prime_j = R_j \sqrt{ \frac{m_j\omega_c}{\hbar}}, \
P^\prime_j =  \frac{P_j}{\sqrt{m_j E_c}}, \\&&
R^\prime_1 = R_1 \sqrt{ \frac{m_1 \omega_c}{\hbar}}, \
P^\prime_1 =  \frac{P_1}{\sqrt{m_1 E_c}}, \\&&
P^\prime_{\eta_1} =  \frac{P_{\eta_1}}{\sqrt{m_{\eta_1} E_c}}, \
P^\prime_{\eta_2} =  \frac{P_{\eta_2}}{\sqrt{m_{\eta_2} E_c}},\\&&
t^\prime =  \omega_c t
, \
H^\prime = \frac{H}{E_c}
, \
T^\prime = \frac{k_B T}{E_c},
\ea
and
\ba
&&
\tilde{\omega}^\prime(t') = \frac{\tilde{\omega}(t)}{\omega_c} 
= \omega_0^\prime \sin\left(\omega_d^\prime t^\prime\right)
,\\&&
\omega^\prime = \frac{\omega}{\omega_c}, \
\omega_0^\prime = \frac{\omega_0}{\omega_c}, \
\omega_d^\prime = \frac{\omega_d}{\omega_c}, 
\\&&
\bar\omega_0' = \frac{\bar\omega_0}{\omega_c} =
\frac{1}{N} \left[1- \exp{(-\omega_{\rm max}')}\right],
\\&&
\Omega_j' = \frac{\Omega_j}{\omega_c}
= - \ln{\left(1-j \bar\omega_0' \right)} \;, 
\\&&
\omega_{\rm max}' = \frac{\omega_{\rm max}}{\omega_c}, \
\Omega_1^\prime = \frac{\Omega_1}{\omega_c} ,
\ea
and
\be
c_j^\prime = \frac{c_j}{\omega_c\sqrt{m m_j \Omega_j \omega_c}} 
= \sqrt{\frac{\xi \hbar \bar\omega_0' \Omega_j'}{m  \omega_c}}
,
\ee\be
%\\&&
c_1^\prime = \frac{c_1}{\omega_c\sqrt{m m_1 \Omega_1 \omega_c }}
= \sqrt{\frac{\xi \hbar \bar\omega_0' \Omega_1'}{m  \omega_c}}
,
\ee
where $E_c = \hbar \omega_c$ and the indices $i$ and $j$
run from 1 to 2 and from 1 to $N$, respectively.
Also, the Hamiltonian $H$ and temperature $T$ can carry any subscripts as
prescribed by a model.
Unless stated otherwise, all these definitions are valid throughout the paper.
Using them, the main equations of the above-mentioned three models can be
written as follows.

The dimensionless Hamiltonian of the model in Eq.~(\ref{Ham_2osc})
takes the form
\be
H_{\rm S}^\prime = \frac{p^{\prime 2}_1}{2} 
+ \frac{p^{\prime 2}_2}{2}
+ \frac{\omega^{\prime 2}}{2} 
\left(q_1^{\prime 2} + q_2^{\prime 2} \right)
+ \frac{\tilde{\omega}^{\prime 2}(t)}{2}
\left( q_2^\prime - q_1^\prime \right)^2 
%s \;
,
\ee
and the equations of motion for the dimensionless
phase-space coordinates
$\left(q_1^\prime, q_2^\prime, p_1^\prime, p_2^\prime\right)$
can be written as
\begin{eqnarray}
\frac{d q_1^\prime}{d t^\prime} &=& p_1^\prime , \\
\frac{d q_2^\prime}{d t^\prime} &=& p_2^\prime ,\\
\frac{d p_1^\prime}{d t^\prime} &=& -\omega^{\prime 2}q_1^\prime 
+ \tilde{\omega}^{\prime 2}(t^\prime)\left(q_2^\prime - q_1^\prime\right) ,\\
\frac{d p_2^\prime}{d t^\prime} &=& -\omega^{\prime 2}q_2^\prime 
- \tilde{\omega}^{\prime 2}(t^\prime)\left(q_2^\prime - q_1^\prime\right)
.
\end{eqnarray}
As long as for this model the value $\omega_c$ does not appear in the Hamiltonian or elsewhere,
for actual computations one could naturally assume $\omega_c \equiv \kappa \omega_d$, where $\kappa$
is any natural number, for example.

The dimensionless Hamiltonian of the model defined in 
Eqs. (\ref{NOSC_ham}-\ref{eq:HSB}) can be written as
\begin{eqnarray}
H_{\rm NB}^\prime &=& \frac{p_1^{\prime 2}}{2} + \frac{p_2^{\prime 2}}{2} +
\frac{\omega^{\prime 2}}{2} q_1^{\prime 2} + 
\frac{\omega^{\prime 2}}{2} q_2^{\prime 2}
\nn\\&+&
\frac{\tilde{\omega}^{\prime 2}(t)}{2}\left(q_2^{\prime2}-q_1^{\prime2}\right)
+ \sum_{j=1}^N \left(\frac{P_j^{\prime 2}}{2} 
+ \frac{1}{2}\Omega_j^{\prime 2} R_j^{\prime 2}\right) \nonumber \\
& - &\left(q_1^\prime+q_2^\prime\right)\sum_{j=1}^N c_j^\prime R_j^\prime ,
\end{eqnarray}
and the equations of motion for the dimensionless phase-space
coordinates $\left(q_1^\prime,q_2^\prime,R_j^\prime,p_1^\prime,p_2^\prime,P_j^\prime
\right)$
are 

\begin{eqnarray}
\frac{d  q_1^\prime}{dt^\prime} &=& p_1^\prime , \\
\frac{d  q_2^\prime}{dt^\prime} &=& p_2^\prime ,\\
\frac{d  R_j^\prime}{dt^\prime} &=& P_j^\prime ,\\
\frac{d  p_1^\prime}{dt^\prime} &=& -\omega^{\prime 2}q_1^\prime
+\tilde{\omega}^{\prime2}(t^\prime)\left(q_2^\prime-q_1^\prime\right) 
+ \sum_{j=1}^Nc_j^\prime R_j^\prime , 
\ea
\ba
\frac{d  p_2^\prime}{dt^\prime} &=& -\omega^{\prime 2}q_2^\prime 
-\tilde{\omega}^{\prime 2}(t)\left(q_2^\prime-q_1^\prime\right) 
+ \sum_{j=1}^Nc_j^\prime R_j^\prime ,  \\
\frac{d P_j^\prime}{dt^\prime} &=& - \Omega^{\prime 2}_j R_j^\prime 
+ c_j^\prime\left(q_1^\prime + q_2^\prime\right)
.
\end{eqnarray}
Finally, the dimensionless Hamiltonian of the NHC model (\ref{NHC_ham})
can be written as
\begin{eqnarray}
H_{\rm NHC}^\prime &=& \frac{p_1^{\prime 2}}{2} + \frac{p_2^{\prime 2}}{2} 
+ \frac{\omega^{\prime 2}}{2}q_1^{\prime 2} + \frac{\omega^{\prime 2}}{2}q_2^{\prime 2}
\nn\\&+&
\frac{\tilde{\omega}^{\prime 2}(t)}{2}\left(q_2^\prime - q_1^\prime\right)^2 
+ \frac{P_1^{\prime 2}}{2} + \frac{\omega_1^{\prime 2}}{2}R_1^{\prime 2}
\nonumber\\
&-&c_1^\prime R_1^\prime \left(q_1^\prime + q_2^\prime \right) 
+ \frac{P_{\eta_1}^{\prime 2}}{2} + \frac{P_{\eta_2}^{\prime 2}}{2} 
\nn\\&+&
gT_{\rm ext}^\prime\eta_1 + T_{\rm ext}^\prime\eta_2 
,
\end{eqnarray}
and the equations of motion for the dimensionless phase-space
coordinates $\left(q_1^\prime,q_2^\prime,R_1^\prime, \eta_1,\eta_2, p_1^\prime,p_2^\prime, P_1^\prime,P_{\eta_1}^\prime,
P_{ \eta_2}^\prime \right)$
are 
\begin{eqnarray}
\frac{d q_1^\prime}{dt^\prime} &=& p_1^\prime \;,\\
\frac{d q_2^\prime}{dt^\prime} &=& p_2^\prime \;,
\\
\frac{d R_1^\prime}{dt^\prime} &=& P_1^\prime \;,
\\
%\ea\ba
\frac{d \eta_1}{dt^\prime} &=& P_{\eta_1}^\prime \;,\\
\frac{d \eta_2}{dt^\prime} &=& P_{\eta_2}^\prime \;,
\\
\frac{d p_1^\prime}{dt^\prime} &=& -\omega^{\prime 2}q_1^\prime
+\tilde{\omega}^{\prime2}(t^\prime)\left(q_2^\prime-q_1^\prime\right)
+ c_1^\prime R_1^\prime \;,\\
\frac{d p_2^\prime}{dt^\prime} &=& -\omega^{\prime 2}q_2^\prime 
-\tilde{\omega}^{\prime 2}(t)\left(q_2^\prime-q_1^\prime\right) 
+ c_1^\prime R_1^\prime \;,
\\
\frac{d P_1^\prime}{dt^\prime} &=& - \Omega^{\prime 2}_jR_1^\prime 
+ c_1^\prime\left(q_1^\prime + q_2^\prime\right) \;,\\
\frac{d P_{\eta_1}^\prime}{dt^\prime} &=& \left(P_1^{\prime 2}-gT_{\rm ext}^\prime\right)
-P_{\eta_1}^\prime P_{\eta_2}^\prime\;,\\
\frac{d P_{\eta_2}^\prime}{dt^\prime} &=& P_{\eta_1}^{\prime 2}
-T_{\rm ext}^\prime \;.
\end{eqnarray}

%%%%%%%%%%%%%%%%%%%%%%%%%%%%%%%%%%%%%%%%%%%%%%%%%%


\begin{thebibliography}{}

%1
\bibitem{whatislife}
E. Schr\"odinger, What is Life? Cambridge University Press,
Cambridge (2013).

%2
\bibitem{qinfo}
M. A. Chang and M. Nielsen, Quantum Computation and Quantum Information.
Cambridge University Press, Cambridge (2011).

%3
\bibitem{engel}
G. S. Engel, \textit{et al.},
%T. R. Calhoun, E. L. Read, T.-K. Ahn, T. Man\u cal, Y.-C. Cheng, R. E. Blankenship, and G. R. Fleming,
%Evidence for wavelike energy transfer through quantum coherence
%in photosynthetic systems
Nature {\bf 446}, 782-786 (2007).

%4
\bibitem{collini}
E. Collini, \textit{et al.},
%C. Y. Wong, K. E. Wilk, P. M. G. Curmi, P. Brumer, and G. D. Scholes,
%Coherently wired light-harvesting in photosynthetic marine algae
%at ambient temperature,
Nature {\bf 463}, 644-647 (2010).

%5
\bibitem{pani}
G. Panitchayangkoon, \textit{et al.},
%D. V. Voronine, D. Abramavicius, J. R. Caram, N. H. C. Lewis, S. Mukamel, and G. S. Engel, 
%Direct Evidence of Quantum Transport in Photosynthetic 
%Light-harvesting Complexes 
Proc. Natl Acad. Sci. {\bf 108}, 20908-20912 (2011).

%6
\bibitem{fle11}
G. R. Fleming, S. F. Huelga, and M. B. Plenio,
%Focus on quantum effects and noise in biomolecules
New J. Phys. {\bf 13}, 115002 pp. 5 (2011).

%7
\bibitem{bri13}
H. J. Briegel and S. Popescu,
%Intra-molecular refrigeration in enzymes
Proc. R. Soc. A {\bf 469}, 20110290 pp. 9 (2013).

%8
\bibitem{poc09}
N. Poccia, A. Ricci, D. Innocenti, and A. Bianconi,
%A possible mechanism for evading temperature quantum decoherence in
%living matter by Feshbach resonance.
Int. J. Molec. Sci. {\bf 10}, 2084-2106 (2009).

%9
\bibitem{vbb97}
A. Valletta, \textit{et al.},
%G. Bardelloni, M. Brunelli, A. Lanzara, A. Bianconi, and N. L. Saini,
%“Tc amplification and pseudogap at a shape resonance in a superlattice of quantum stripes” ” 
J. Superconductivity {\bf 10}, 383-387 (1997).

%10
\bibitem{chi10}
A. W. Chin, A. Datta, F. Caruso, S. F. Huelga, and M. B. Plenio,
%Noise-assisted energy transfer in quantum networks and light-harvesting complexes. 
New J. Phys. {\bf 12}, 065002 pp. 16 (2010).

%11
\bibitem{EntangHighT}
F. Galve, L. A. Pach\'on and D. Zueco, 
Phys. Rev. Lett. {\bf 105}, 180501 pp. 4 (2010).

%12
\bibitem{EntangHighT2}
F. Galve, G. L. Giorgi, and R. Zambrini,
%Entanglement dynamics of nonidentical oscillators under decohering environments
Phys. Rev. A {\bf 81}, 062117 pp. 10 (2010).

%13
\bibitem{gue12}
G. G. Guerreschi, J. Cai, S. Popescu, H. J. Briegel,
%Persistent dynamic entanglement from classical motion: 
%how bio-molecular machines can generate nontrivial quantum states
New J. Phys. {\bf 14}, 053043 pp. 21 (2012).

%14
\bibitem{pachon}
A. F. Estrada and L. A. Pachon,
%Quantum Limit for Driven Linear Non-Markovian Open-Quantum-Systems,
arXiv:1411.3382 [quant-ph] (2014).

%15
\bibitem{amico}
L. Amico, R. Fazio, A. Osterloh, and V. Vedral,
Rev. Mod. Phys. {\bf 80}, 517-576 (2008).

%16
\bibitem{horodecki}
R. Horodecki, P. Horodecki, M. Horodecki, and K. Horodecki,
Rev. Mod. Phys. {\bf 81}, 865-942 (2009).

%17
\bibitem{lecture_notes}
Irreversible Quantum Dynamics, Lecture Notes in Physics,
F. Benatti and R. Floreanini eds. 
(Springer, Berlin, 2013).

%18
\bibitem{weiss}
U. Weiss, Quantum Dissipative Systems (World Scientific, Singapore, 2008).

%19
\bibitem{wigner}
E. Wigner, Phys. Rev. {\bf 40}, 749-759 (1932).

%20
\bibitem{DistFunctions}
M. Hillery, R. F. O'Connell, M. O. Scully and E. P. Wigner,
Phys. Rep. {\bf 106}, 121-167 (1984).

%21 
\bibitem{lee}
H. Lee, Phys. Rep. {\bf 259}, 147-211 (1995).

%22
\bibitem{Liquids}
M. P. Allen and D. J. Tildesley,
Computer Simulation of Liquids. Clarendon Press, Oxford (1989).

%23
\bibitem{UnderstandMolSim}
D. Frenkel and B. Smit,
Understanding Molecular Simulation. Academic Press, San Diego (2002).

%24
\bibitem{QOptics}
C. C. Gerry and P. L. Knight,
Introductory quantum optics. Cambridge University Press, Cambridge (2005).

%25
\bibitem{quantumnoise}
C. W. Gardiner and P. Zoller,
Quantum Noise. Springer, Berlin (2004).

%26
\bibitem{nhc}
G. J. Martyna, M. L. Klein and M. Tuckerman,
J. Chem. Phys. {\bf 97}, 2635-2643 (1992).

%27
\bibitem{deco1}
P. Zanardi and M. Rasetti, Phys. Rev. Lett. {\bf 79}, 3306 (1997).

%28
\bibitem{deco2}
D. A. Lidar, I. L. Chuang, and K. B. Whaley, 
Phys. Rev. Lett. {\bf 81}, 2594 (1998).

%29
\bibitem{deco3}
L.-M. Duan and G.-C. Guo, Phys. Rev. A {\bf 57}, 737 (1998).

%30
\bibitem{ilya}
A. Sergi, I. Sinayskiy, and F. Petruccione,
%``Numerical and Analytical Approach to the Quantum Dynamics
%of Two Coupled Spins in Bosonic Baths'',
Phys. Rev. A {\bf 80}, 012108 pp. 7 (2009).

%31
\bibitem{phonon_sqz}
S. L. Johnson, \textit{et al.},
%P. Beaud, E. Vorobeva, C. J. Milne, \'E. D. Murray, S. Fahy, and G. Ingold, 
Phys. Rev. Lett. {\bf 102}, 175503 pp. 4 (2009).

%32
\bibitem{condense_squeeze}
S.-L. Ma, P.-B. Li, A.-P. Fang, S.-Y. Gao, and F.-L. Li, 
Phys. Rev. A {\bf 88}, 013837 pp. 5 (2013).

%33
\bibitem{condense_squeeze2}
T. Altanhan and B. S. Kandemir,
J. Phys.: Condens. Matter {\bf 5}, 6729-6736 (1993).

%34
\bibitem{spec_squeeze}
D. J. Wineland, J. J. Bollinger, W. M. Itano, and D. J. Heinzen,
Phys. Rev. A {\bf 50}, 67-88 (1994).

%35
\bibitem{qi_squeeze}
V. C. Usenko and R. Filip, 
New J. Phys. {\bf 13}, 113007 pp. 14 (2011).

%36
\bibitem{grav_squeeze} 
S. Dwyer, \textit{et al.},
%L. Barsotti, S. S. Y. Chua, M. Evans, M. Factourovich, D. Gustafson, T. Isogai, K. Kawabe, A. Khalaidovski, P. K. Lam, M. Landry, N. Mavalvala, D. E. McClelland, G. D. Meadors, C. M. Mow-Lowry, R. Schnabel, R. M. S. Schofield, N. Smith-Lefebvre, M. Stefszky, C. Vorvick and D. Sigg,
Opt. Express {\bf 21}, 19047-19060 (2013).

%37
\bibitem{odor}
J. C. Brookes, F. Hartoutsiou, A. P. Horsfield, and A. M. Stoneham,
%Could Human Recognize Odo by Phonon Assisted Tunneling?
Phys. Rev. Lett. {\bf 98}, 038101 (2007).

%38
\bibitem{QphysVlad}
V. Zelevinsky, Quantum Physics; Vol. I. Wiley-VCH, Weinhein (2011).

%39
\bibitem{Ballentine}
L. E. Ballentine, Quantum mechanics. World Scientific, Amsterdam (2005).

%40
\bibitem{nonHamTherm}
A. Sergi and F. Petruccione, J. Phys. A {\bf 41}, 355304 pp. 14 (2008).

%41
\bibitem{Moyal_paper}
J. E. Moyal, Proc. Camb. Philos. Soc. {\bf 45}, 99-124 (1949).

%42
\bibitem{nose}
S. Nos\'e, Mol. Phys. {\bf 52}, 255-268 (1984).

%43
\bibitem{hoover}
W. G. Hoover, Phys. Rev. A {\bf 31}, 1695-1697 (1985).

%44
\bibitem{nheom}
A. Sergi and M. Ferrario, Phys. Rev. E. {\bf 64}, 056125 pp. 9 (2001).

%45 
\bibitem{nh_esm}
A. Sergi, 
%Non-Hamiltonian Equilibrium Statistical Mechanics
Phys. Rev. E. {\bf 67}, 021101 pp. 7 (2003). 

%46
\bibitem{geometry}
A. Sergi and P. V. Giaquinta,
%On the geometry and entropy of non-Hamiltonian phase space,
J. Stat. Mech.  {\bf 02}, P02013 pp. 20 (2007).

%47
\bibitem{linearquantum}
E. A. Martinez and J. P. Paz, Phys. Rev. Lett. {\bf 110}, 130406 pp. 4 (2013).

%48
\bibitem{Floquet}
A. Lindner and H. Freese, J. Phys. A {\bf 27}, 5565-5571 (1994).

%49
\bibitem{leggett}
A. J. Leggett, \textit{et al.},
%S. Chakravarty, A. T. Dorsey, M. P. A. Fisher, A. Garg, and M. Zwerger, 
Rev. Mod. Phys. {\bf 59}, 1-85 (1987).

%50
\bibitem{makri}
N. Makri and K. Thompson, J. Phys. Chem. {\bf 291}, 101-109 (1998).

%51
\bibitem{linear1}
K. Thompson and N. Makri, J. Chem. Phys., {\bf 110}, 1343-1353 (1999).

%52
\bibitem{linear2}
N. Makri, J. Phys. Chem. B, {\bf 103}, 2823-2829 (1999).

%53
\bibitem{dlamini}
N. Dlamini and A. Sergi,
%Quantum dynamics in classical thermal baths,
Comp. Phys. Comm. {\bf 184}, 2474-2477 (2013).

%54
\bibitem{b3}
A. Sergi,
%Deterministic constant-temperature dynamics for dissipative quantum systems,
J. Phys. A {\bf 40}, F347-F354 (2007).

%55
\bibitem{reversibleintegrators}
G. J. Martyna, M. E. Tuckerman, D. J. Tobias and  M. L. Klein,
Mol. Phys. {\bf 87}, 1117-1157 (1996).

%56
\bibitem{reverse_int}
A. Sergi, M. Ferrario and D. Costa, Molec. Phys. {\bf 97}, 825-832 (1999).

%57
\bibitem{yoshida}
H. Yoshida, Phys. Lett. A {\bf 150}, 262-268 (1990).

%58
\bibitem{Goldstein}
H. Goldstein, Classical Mechanics. Addison-Wesley, Reading MA (1980).

%59
\bibitem{schleich}
W. P. Schleich, Quantum Optics in Phase Space. Wiley, Berlin (2001).

%60
\bibitem{zhang}
H. Zhang, K. Siegrist, D. F. Plusquellic and S. K. Gregurick, 
J. Am. Chem. Soc. {\bf 130}, 17846-17857 (2008).

\end{thebibliography}
\end{document}